\documentclass{article}

\usepackage{arxiv}

\usepackage[utf8]{inputenc} 
\usepackage[T1]{fontenc}    
\usepackage{hyperref}       
\usepackage{url}            
\usepackage{booktabs}       
\usepackage{amsfonts}       
\usepackage{nicefrac}       
\usepackage{microtype}      
\usepackage{lipsum}		
\usepackage{graphicx}
\usepackage{natbib}
\usepackage{doi}

\title{\textit{Filters of Identity}: AR Beauty and the Algorithmic Politics of the Digital Body}

\author{ \href{https://orcid.org/0000-0003-2523-6901}{\includegraphics[scale=0.06]{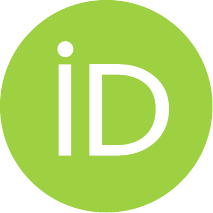}\hspace{1mm}Miriam Doh} \\
	ISIA Lab - Université de Mons\\
	IRIDIA Lab - Université Libre de Bruxelles\\
	Belgium\\
	\\
	\And
	\href{https://orcid.org/0009-0009-1511-3925}{\includegraphics[scale=0.06]{orcid.pdf}\hspace{1mm}Corinna Canali} \\
	Design, Diversity, and New Commons Research Group, \\Universität der Künste Berlin \\ Weizenbaum Institute\\
	Germany \\
	\And
	\href{https://orcid.org/0000-0001-5985-691X}{\includegraphics[scale=0.06]{orcid.pdf}\hspace{1mm}Nuria Oliver} \\
	Ellis Alicante \\Spain \\ 
}

\date{}

\renewcommand{\headeright}{}
\renewcommand{\undertitle}{}
\renewcommand{\shorttitle}{\textit{Filters of Identity}: AR Beauty and the Algorithmic Politics of the Digital Body}

\hypersetup{
pdftitle={\textit{Filters of Identity}: AR Beauty and the Algorithmic Politics of the Digital Body},
pdfsubject={q-bio.NC, q-bio.QM},
pdfauthor={David S.~Hippocampus, Elias D.~Striatum},
pdfkeywords={First keyword, Second keyword, More},
}

\begin{document}
\maketitle

\begin{abstract}
	This position paper situates AR beauty filters within the broader debate on Body Politics in HCI. We argue that these filters are not neutral tools but technologies of governance that reinforce racialized, gendered, and ableist beauty standards. Through naming conventions, algorithmic bias, and platform governance, they impose aesthetic norms while concealing their influence. To address these challenges, we advocate for transparency-driven interventions and a critical rethinking of algorithmic aesthetics and digital embodiment.
\end{abstract}


\section*{Note}
This work was presented at the ``Body Politics:
Unpacking Tensions and Future Perspectives For Body-Centric Design Research in HCI" workshop at the ACM (Association for Computing Machinery) CHI conference on Human Factors in Computing Systems 2025.
\section{Introduction}
AR beauty filters are marketed as tools for self-expression, yet they subtly enforce racialized, gendered, and ableist beauty standards through opaque algorithmic modifications. Rather than simply expanding self-representation, these technologies, along with content moderation algorithms on social platforms, contribute to the construction of aesthetic hierarchies. While beauty filters impose specific visual norms, content moderation systems reinforce these standards by determining which content is amplified or suppressed, together shaping a stratified landscape of digital desirability.

Situating beauty filters within Body Politics in HCI, this position paper argues that these tools do more than mediate appearance—they function as technologies of governance, shaping visibility, desirability, and exclusion in digital spaces. Building on our previous research, we explore how filter naming conventions, algorithmic biases, and platform moderation policies produce and sustain hegemonic beauty norms, often contradicting platform diversity narratives.

Addressing the themes of this workshop, we engage with the three key questions proposed:
\textit{(Q1) What theories have influenced body-centric research in HCI, and how have bodies been conceptualized?
(Q2) What new openings or ethical pitfalls emerge when attending to marginalized or more-than-human bodies through different research methods?
(Q3) How can we account for the multiplicity of bodies while preserving the significance of individual experiences in body-centric research?}

By framing beauty filters within the politics of algorithmic aesthetics, this paper aims to spark a critical discussion on transparency, customization, and platform accountability in digital beauty governance.

\subsection{\textbf{(Q1)} Beauty Filters and the Politics of Digital Embodiment}

Body-centric HCI studies conceptualize the gendered body as a site of negotiation between digital productivity and sociocultural norms, aligning with feminist critiques of capitalism, technology, and the body (\textit{e.g.}, \cite{federici2004caliban,haraway2013cyborg}). Within this framework, AR beauty filters emerge as more than mere appearance enhancements \cite{isakowitsch2022augmented,cug2022beauty,fribourg2021mirror}. They work as powerful technologies of gender \cite{de1987technologies}, encoding and enforcing prescriptive beauty ideals \cite{Leeat_Shnabel_Glick_2019} through algorithmic design, benefiting platforms while shifting responsibility onto users.

Objectification Theory \cite{Fredrickson_Roberts_1997} provides a key lens for understanding these dynamics, showing how individuals—particularly women—internalize an external gaze, leading to self-surveillance and body dissatisfaction. AR beauty filters amplify this process by embedding dominant beauty standards into their algorithms, subtly dictating how users modify their digital self-representation under the guise of choice and empowerment. This disciplining effect extends to filter naming conventions, which, rather than describing their actual effects, actively shape expectations of digital beauty. Our analysis \cite{Doh_Canali_Karagianni_2024} of TikTok filters shows that names such as ``Princess Makeup," ``Pure Eyes," and ``Prettiest" reinforce narrow, feminized beauty ideals, linking attractiveness to idealized youth, innocence, and purity. The linguistic framing of filters constructs a hierarchy of desirability, associating smooth skin, symmetry, and delicate features with idealized beauty while obscuring the extent of algorithmic facial modifications.\footnote{For instance, the ``Prettiest" filter, with 62.2M posts, was used by 98.66\% female users, reinforcing how language encodes aesthetic norms and dictates digital desirability in gendered ways.}

Yet the influence of AR beauty filters extends beyond individual aesthetics to platform-wide governance, where content moderation algorithms and policies dictate whose bodies are visible and desirable. 
Our research into TikTok’s AR filter ecosystem reveals that the platform not only develops and promotes beauty filters shaped by intersectional biases but also actively curates (in)visibility by incentivizing filter use \cite{doh_tiktok}. Prior investigations \cite{Biddle_Ribeiro_Dias_2020,Hern_2020} have exposed internal moderation documents detailing exclusionary content suppression policies, banning users deemed ``too ugly, poor, or disabled" from appearing on the \textit{For You} feed. The suppression criteria included ``abnormal body shape," ``ugly facial features," ``too many wrinkles,' or ``shabby environments." These practices illustrate how the politics of digital embodiment in HCI extend beyond individual filters to platform-wide visibility infrastructures. AR beauty filters function within a broader system of algorithmic governance, shaping who gets seen, how they are represented, and which bodies are deemed desirable in digital spaces.

\subsubsection{Postfeminist and Neoliberal Frameworks of Beauty Optimization}
Within their inherent gender bias, AR beauty filters operate in a postfeminist media culture \cite{gill2007gender, mcrobbie2004post}, where self-surveillance is reframed as empowerment, and normative self-improvement becomes an obligation rather than a choice. Through AI-driven enhancements and opaque platform governance, beauty filters seamlessly integrate into this logic, offering users the illusion of control over their digital self-presentation while reinforcing deeply embedded gendered norms of attractiveness. Under the neoliberal, postfeminist mandate of self-optimization \cite{gill2007gender,mcrobbie2004post,elias2017aesthetic,hearn2020beguiling} body-computer interactions become a site of real-time governmentality \footnote{Governmentality, as formulated by Michel Foucault, refers to the governance of people through positive means rather than coercion, shaping their conduct by fostering willing participation. Neoliberal governance exemplifies this by both isolating individuals as self-responsible entrepreneurial units and binding them to collective economic and political structures \cite{brown2015undoing}}, of ``conduct of conduct" \cite{foucault2007security}  through rhetoric of autonomy, self-actualization, and self-realization \cite{Madsen2014}, while the body  turns into a profitable site of aesthetic \cite{elias2017aesthetic} and glamour labor \cite{wissinger2015year,hearn2020beguiling}––\textit{i.e.}, gendered practices that emphasize female attractiveness ``at whatever cost to the self" \cite{elias2017aesthetic}. Within platform economies, the idealized self is produced and circulated as a tool for market-driven interactions. AR beauty filters thus function as a mode of gendered governmentality. By conditioning individuals to internalize platform-driven beauty standards, these technologies extend neoliberal postfeminist surveillance, where the feminized body is simultaneously hyper-visible and invisibilized, optimized for digital desirability yet interchangeable within algorithmic economies of attention.
\begin{figure}[ht]
    \centering
    \includegraphics[width=0.8\linewidth]{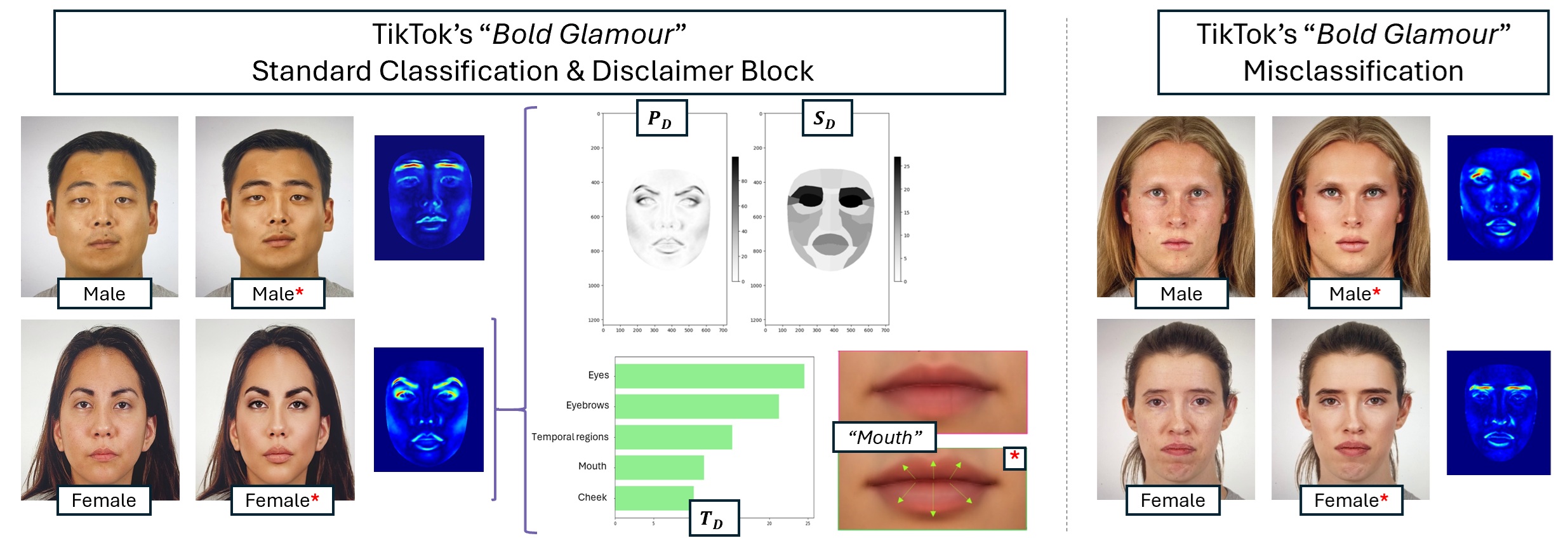}
    \caption{Analysis of TikTok's Bold Glamour filter transformations from \cite{doh_tiktok}. (Left) Example outputs of the filter applied to male and female faces (``*" indicates filtered images). (Center) \texttt{Disclaimer Block} visualizations: $P_D$ highlights pixel-level changes, $S_D$ represents semantic mask transformations, and $T_D$ quantifies intensity variations. A zoom-in on the `Mouth' region reveals how the filter enlarges lips. (Right) Two cases of misclassification: the first row shows a male face receiving female-targeted transformations, while the second row shows a female face with a misclassified male-targeted modification. Faces from \cite{ma2015cfd}}
    \label{fig:boldglamour}
\end{figure}

\subsection{\textbf{(Q2)} Algorithmic Bias and Aesthetic Exclusion in AR Filters}

Despite their potential for inclusivity, current implementations of AR beauty filters reinforce exclusion and uniformity over diversity (\textit{e.g.}, \cite{riccio2022openfilter} and \cite{doh_tiktok}). Rather than enabling true self-expression, they function within a governance system that dictates aesthetic priorities and modifications. By embedding racial, gender, and algorithmic biases (\textit{e.g.}, \cite{riccio2024mirror} and \cite{doh_tiktok}), these technologies shape digital desirability through opaque aesthetic norms.

A remarkable example is TikTok’s \textit{Bold Glamour} filter, Fig. \ref{fig:boldglamour}, which employs hyperrealistic facial modifications that seamlessly integrate into users’ faces, making algorithmic interventions nearly imperceptible \cite{Ruggeri_2023}. 
In fact, unlike traditional AR filters that apply visible overlays, \textit{Bold Glamour} modifies one's appearance in ways that can be difficult to detect, doing so while privileging Eurocentric and cisgender facial structures that systematically exclude non-conforming bodies. This reflects a broader trend in technological design, where digital systems tend to reinforce normative assumptions about bodies, making those that deviate from dominant beauty standards—such as disabled, fat, Indigenous, and Black individuals—less visible or misrepresented \cite{spiel2021bodies}.

These biased algorithmic mechanisms manifest in three key ways, as identified in our study of \textit{Bold Glamour} \cite{doh_tiktok}
:  
\textbf{(1) Facial recognition bias:} \textit{Bold Glamour} applies different transformations based on algorithmically inferred gender, assuming binary male/female classifications without user control. It also exhibits higher misclassification rates for certain racial groups, especially Black women, leading to aesthetic distortions and exclusionary outputs.  
\textbf{(2) Enforced aesthetic conformity:} Our dataset analysis shows that \textit{Bold Glamour} systematically aligns facial features toward Western beauty standards, including lighter skin tones, fuller lips, smaller noses, and larger eyes. These alterations reinforce hegemonic beauty ideals, contradicting the platform's stated commitment to diversity in filter creation/circulation.  
\textbf{(3) Lack of agency in digital self-representation:} Users often engage with beauty filters without full awareness of the extent of their modifications, potentially leading to distorted self-perceptions and increased self-objectification. This kind of algorithmic curation reinforces societal hierarchies of desirability and acceptability, making certain bodies more valuable than others under profit-driven platform governance \cite{gillespie2017governance}. Hence, if, as Winch \cite{winch2015brand} notes, managing the body is the ``means by which women acquire and display their cultural capital," the tools for content production and curation increasingly intersect self-surveillance, platform surveillance \cite{zuboff2023age}, and the commodification of the feminized body.

Beyond algorithmic biases, a key ethical concern is the gap between platform policies and actual practices. TikTok’s Effect House explicitly prohibits filters that promote unattainable beauty ideals, yet the platform continues to develop and promote AR beauty filters that subtly reinforce them.

This contradiction highlights how platform governance conceals aesthetic biases under the guise of inclusivity. Addressing this ethical pitfall requires not only questioning filter design but also critically examining the regulatory contradictions that sustain these biases.

AI-driven beauty filters do not merely reflect beauty norms; they actively construct and reinforce them, marginalizing non-normative bodies while generating platform profit. The increasing integration of AI-generated faces \cite{doh2024my,doh2025algorithmsplayfavoriteslookism} and synthetic beauty filters raises critical questions about digital embodiment. \textit{What happens when beauty is no longer determined by human perception but is instead produced by algorithms? How do these technologies reshape identity and authenticity?} These concerns are particularly relevant in more-than-human HCI, where algorithmic beauty norms increasingly override human-defined aesthetic representations.  

\subsection{\textbf{(Q3)} Ensuring Multiplicity in Digital Beauty - Transparency and Individuality in Beauty Filters}

A key challenge in body-centric HCI research is ensuring that digital beauty technologies embrace bodily diversity while allowing individuals to retain agency over their self-representation. Rather than reinforcing a singular aesthetic ideal, these systems should provide space for multiple forms of beauty to coexist, reflecting cultural and personal differences. However, current AR beauty filters often prioritize uniformity over customization, limiting users’ ability to shape their digital identities in authentic ways. Addressing this issue requires interventions that enhance \textbf{(1) Transparency and (2) Personalization}.
Our previous work \cite{Doh_Canali_Karagianni_2024} introduced the \texttt{Disclaimer Block}, Fig. \ref{fig:boldglamour}, a tool designed to expose the algorithmic transformations applied by beauty filters. By making these modifications explicit, the \texttt{Disclaimer Block} ensures that users can critically engage with their digital self-representation rather than passively conforming to hidden aesthetic norms. 

However, achieving true diversity in digital aesthetics requires not only customizable filters but also a shift in content moderation policies. Platforms should rethink how their moderation algorithms amplify or suppress different beauty standards, ensuring that diverse representations are not algorithmically marginalized in favor of a singular, market-driven ideal.
Future beauty filters should prioritize adaptive, user-driven modifications over fixed, one-size-fits-all enhancements: \textbf{(2.1) Manual intensity controls: }Allow users to selectively adjust filter effects rather than enforcing predefined transformations. \textbf{(2.2) Disabling automatic gender classification: }For some filter, like \textit{Bold Glamour}, instead of applying modification based on inferred identity, filters should enable users to define their aesthetic preferences. \textbf{(2.3) Diversity beauty presets:} Instead of reflecting a singular beauty model, filters should provide multiple aesthetic templates, drawing from diverse cultural and individual beauty norms.

By shifting from passive algorithmic enforcement to active, self-determined representation, beauty filters can become tools of empowerment rather than normative regulation. This shift ensures that multiplicity is not just acknowledged, but actively preserved in digital beauty technologies.

\section{Conclusion and Next Steps}

This position paper examines AR beauty filters as tools of algorithmic governance that shape digital desirability, inclusion, and self-representation. Through the case of TikTok’s Bold Glamour and our previous research, we have shown how these filters encode racialized, gendered, and ableist norms under the guise of personalization, reinforcing power structures in digital embodiment. While our discussion of the Disclaimer Block illustrates one approach to algorithmic transparency, transparency alone is insufficient.

Moving forward, we invite workshop participants to explore:
(1) \textit{Can user-driven customization counteract algorithmic beauty enforcement?}
(2) \textit{What alternatives to transparency can disrupt hidden aesthetic hierarchies?}
(3) \textit{How can research promote regulatory or design interventions to ensure platform accountability?}

Rather than concluding with a fixed answer, we see this position paper as an invitation to critically question the intersection of AI, embodiment, and digital beauty within body-centric HCI. By centering multiplicity and self-determination, we aim to contribute to a broader conversation on how beauty technologies might be reimagined—not as silent enforcers of desirability but as tools for diverse and self-directed digital embodiment.

\section*{Acknowledgements}
M.D. was supported by the ARIAC project (No. 2010235), funded by the Service Public de Wallonie (SPW Recherche) and FNRS (National Fund for Scientific Research).

C.C. was funded by Berlin University of the Arts and Weizenbaum Institute for the Networked Society, Berlin, funded by the German Federal Ministry of Education and Research (BMBF). 
N.O. has been partially funded by a nominal grant received at the
ELLIS Unit Alicante Foundation from the Regional Government of Valencia in
Spain (Convenio Singular signed with Generalitat Valenciana, Conselleria de Innovación, Industria,
Comercio y Turismo, Dirección General de Innovación) and by the European Union EU - HE ELIAS – Grant Agreement 101120237.
Views and opinions expressed are however those of the author(s) only and do not necessarily reflect
those of the European Union or the European Health and Digital Executive Agency (HaDEA).

\bibliographystyle{ACM-Reference-Format}
\bibliography{references}


\begin{thebibliography}{30}


\ifx \showCODEN    \undefined \def \showCODEN     #1{\unskip}     \fi
\ifx \showDOI      \undefined \def \showDOI       #1{#1}\fi
\ifx \showISBNx    \undefined \def \showISBNx     #1{\unskip}     \fi
\ifx \showISBNxiii \undefined \def \showISBNxiii  #1{\unskip}     \fi
\ifx \showISSN     \undefined \def \showISSN      #1{\unskip}     \fi
\ifx \showLCCN     \undefined \def \showLCCN      #1{\unskip}     \fi
\ifx \shownote     \undefined \def \shownote      #1{#1}          \fi
\ifx \showarticletitle \undefined \def \showarticletitle #1{#1}   \fi
\ifx \showURL      \undefined \def \showURL       {\relax}        \fi
\providecommand\bibfield[2]{#2}
\providecommand\bibinfo[2]{#2}
\providecommand\natexlab[1]{#1}
\providecommand\showeprint[2][]{arXiv:#2}

\bibitem[Biddle et~al\mbox{.}(2020)]%
        {Biddle_Ribeiro_Dias_2020}
\bibfield{author}{\bibinfo{person}{S. Biddle}, \bibinfo{person}{P.~V. Ribeiro}, {and} \bibinfo{person}{T. Dias}.} \bibinfo{year}{2020}\natexlab{}.
\newblock \bibinfo{booktitle}{\emph{Invisible Censorship}}.
\newblock The Intercept.
\newblock
\urldef\tempurl%
\url{https://theintercept.com/2020/03/16/tiktok-app-moderators-users-discrimination/}
\showURL{%
\tempurl}
\newblock
\shownote{[Accessed: Nov 2024]}.


\bibitem[Brown(2015)]%
        {brown2015undoing}
\bibfield{author}{\bibinfo{person}{Wendy Brown}.} \bibinfo{year}{2015}\natexlab{}.
\newblock \showarticletitle{Undoing the demos: Neoliberalism's stealth revolution}.
\newblock  (\bibinfo{year}{2015}).
\newblock


\bibitem[Cug et~al\mbox{.}(2022)]%
        {cug2022beauty}
\bibfield{author}{\bibinfo{person}{Juraj Cug}, \bibinfo{person}{Alina T{\u{a}}nase}, \bibinfo{person}{Cristian~Ionu{\c{t}} Stan}, {and} \bibinfo{person}{Tan{\c{t}}a~Camelia Chitc{\u{a}}}.} \bibinfo{year}{2022}\natexlab{}.
\newblock \showarticletitle{Beauty filters for physical attractiveness: Idealized appearance and imagery, visual content and representations, and negative behaviors and sentiments}.
\newblock \bibinfo{journal}{\emph{Journal of Research in Gender Studies}} \bibinfo{volume}{12}, \bibinfo{number}{2} (\bibinfo{year}{2022}), \bibinfo{pages}{33--47}.
\newblock


\bibitem[Doh et~al\mbox{.}(2024)]%
        {Doh_Canali_Karagianni_2024}
\bibfield{author}{\bibinfo{person}{Miriam Doh}, \bibinfo{person}{Corinna Canali}, {and} \bibinfo{person}{Anastasia Karagianni}.} \bibinfo{year}{2024}\natexlab{}.
\newblock \showarticletitle{Pixels of Perfection and Self-Perception: Deconstructing AR Beauty Filters and Their Challenge to Unbiased Body Image}. In \bibinfo{booktitle}{\emph{Proceedings of the 2024 ACM International Conference on Interactive Media Experiences}}. \bibinfo{publisher}{ACM}, \bibinfo{address}{Stockholm, Sweden}, \bibinfo{pages}{349--353}.
\newblock


\bibitem[Doh et~al\mbox{.}(2025a)]%
        {doh_tiktok}
\bibfield{author}{\bibinfo{person}{Miriam Doh}, \bibinfo{person}{Corinna Canali}, {and} \bibinfo{person}{Nuria Oliver}.} \bibinfo{year}{2025}\natexlab{a}.
\newblock \showarticletitle{What TikTok Claims, What Bold Glamour Does: A Filter’s Paradox}. In \bibinfo{booktitle}{\emph{Proceedings of the 2025 ACM Conference on Fairness, Accountability, and Transparency}} \emph{(\bibinfo{series}{FAccT '25})}. \bibinfo{publisher}{Association for Computing Machinery}, \bibinfo{address}{New York, NY, USA}, \bibinfo{pages}{1902–1915}.
\newblock
\showISBNx{9798400714825}
\urldef\tempurl%
\url{https://doi.org/10.1145/3715275.3732126}
\showDOI{\tempurl}


\bibitem[Doh et~al\mbox{.}(2025b)]%
        {doh2025algorithmsplayfavoriteslookism}
\bibfield{author}{\bibinfo{person}{Miriam Doh}, \bibinfo{person}{Aditya Gulati}, \bibinfo{person}{Matei Mancas}, {and} \bibinfo{person}{Nuria Oliver}.} \bibinfo{year}{2025}\natexlab{b}.
\newblock \bibinfo{title}{When Algorithms Play Favorites: Lookism in the Generation and Perception of Faces}.
\newblock
\newblock
\showeprint[arxiv]{2506.11025}~[cs.LG]
\urldef\tempurl%
\url{https://arxiv.org/abs/2506.11025}
\showURL{%
\tempurl}


\bibitem[Doh and Karagianni(2024)]%
        {doh2024my}
\bibfield{author}{\bibinfo{person}{Miriam Doh} {and} \bibinfo{person}{Anastasia Karagianni}.} \bibinfo{year}{2024}\natexlab{}.
\newblock \bibinfo{title}{"My Kind of Woman": Analysing Gender Stereotypes in AI through The Averageness Theory and EU Law}.
\newblock
\newblock
\showeprint[arxiv]{2407.17474}~[cs.CY]
\urldef\tempurl%
\url{https://arxiv.org/abs/2407.17474}
\showURL{%
\tempurl}


\bibitem[Elias et~al\mbox{.}(2017)]%
        {elias2017aesthetic}
\bibfield{author}{\bibinfo{person}{Ana Elias}, \bibinfo{person}{Rosalind Gill}, {and} \bibinfo{person}{Christina Scharff}.} \bibinfo{year}{2017}\natexlab{}.
\newblock \bibinfo{booktitle}{\emph{Aesthetic labour: Beauty politics in neoliberalism}}.
\newblock \bibinfo{publisher}{Springer}.
\newblock


\bibitem[Federici(2004)]%
        {federici2004caliban}
\bibfield{author}{\bibinfo{person}{Silvia Federici}.} \bibinfo{year}{2004}\natexlab{}.
\newblock \bibinfo{booktitle}{\emph{Caliban and the Witch}}.
\newblock \bibinfo{publisher}{Autonomedia}.
\newblock


\bibitem[Foucault(2007)]%
        {foucault2007security}
\bibfield{author}{\bibinfo{person}{Michel Foucault}.} \bibinfo{year}{2007}\natexlab{}.
\newblock \bibinfo{booktitle}{\emph{Security, territory, population: lectures at the Coll{\`e}ge de France, 1977-78}}.
\newblock \bibinfo{publisher}{Springer}.
\newblock


\bibitem[Fredrickson and Roberts(1997)]%
        {Fredrickson_Roberts_1997}
\bibfield{author}{\bibinfo{person}{Barbara~L Fredrickson} {and} \bibinfo{person}{Tomi-Ann Roberts}.} \bibinfo{year}{1997}\natexlab{}.
\newblock \showarticletitle{Objectification theory: Toward understanding women's lived experiences and mental health risks}.
\newblock \bibinfo{journal}{\emph{Psychology of women quarterly}} \bibinfo{volume}{21}, \bibinfo{number}{2} (\bibinfo{year}{1997}), \bibinfo{pages}{173--206}.
\newblock


\bibitem[Fribourg et~al\mbox{.}(2021)]%
        {fribourg2021mirror}
\bibfield{author}{\bibinfo{person}{Rebecca Fribourg}, \bibinfo{person}{Etienne Peillard}, {and} \bibinfo{person}{Rachel Mcdonnell}.} \bibinfo{year}{2021}\natexlab{}.
\newblock \showarticletitle{Mirror, mirror on my phone: Investigating dimensions of self-face perception induced by augmented reality filters}. In \bibinfo{booktitle}{\emph{2021 IEEE International Symposium on Mixed and Augmented Reality (ISMAR)}}. IEEE, \bibinfo{pages}{470--478}.
\newblock


\bibitem[Gill(2007)]%
        {gill2007gender}
\bibfield{author}{\bibinfo{person}{R. Gill}.} \bibinfo{year}{2007}\natexlab{}.
\newblock \bibinfo{booktitle}{\emph{Gender and the Media}}.
\newblock \bibinfo{publisher}{Polity}.
\newblock


\bibitem[Gillespie(2017)]%
        {gillespie2017governance}
\bibfield{author}{\bibinfo{person}{Tarleton Gillespie}.} \bibinfo{year}{2017}\natexlab{}.
\newblock \showarticletitle{Governance of and by platforms}.
\newblock \bibinfo{journal}{\emph{SAGE handbook of social media}} (\bibinfo{year}{2017}), \bibinfo{pages}{254--278}.
\newblock


\bibitem[Haraway(2013)]%
        {haraway2013cyborg}
\bibfield{author}{\bibinfo{person}{Donna Haraway}.} \bibinfo{year}{2013}\natexlab{}.
\newblock \showarticletitle{A cyborg manifesto: Science, technology, and socialist-feminism in the late twentieth century}.
\newblock In \bibinfo{booktitle}{\emph{The transgender studies reader}}. \bibinfo{publisher}{Routledge}, \bibinfo{pages}{103--118}.
\newblock


\bibitem[Hearn and Banet-Weiser(2020)]%
        {hearn2020beguiling}
\bibfield{author}{\bibinfo{person}{Alison Hearn} {and} \bibinfo{person}{Sarah Banet-Weiser}.} \bibinfo{year}{2020}\natexlab{}.
\newblock \showarticletitle{The beguiling: Glamour in/as platformed cultural production}.
\newblock \bibinfo{journal}{\emph{Social Media+ Society}} \bibinfo{volume}{6}, \bibinfo{number}{1} (\bibinfo{year}{2020}), \bibinfo{pages}{2056305119898779}.
\newblock


\bibitem[Hern(2020)]%
        {Hern_2020}
\bibfield{author}{\bibinfo{person}{A. Hern}.} \bibinfo{year}{2020}\natexlab{}.
\newblock \showarticletitle{TikTok “tried to filter out videos from ugly, poor or disabled users”}.
\newblock \bibinfo{journal}{\emph{The Guardian}} (\bibinfo{year}{2020}).
\newblock
\urldef\tempurl%
\url{https://tinyl.io/BjRa}
\showURL{%
\tempurl}
\newblock
\shownote{[Accessed: Nov 2024]}.


\bibitem[Isakowitsch(2022)]%
        {isakowitsch2022augmented}
\bibfield{author}{\bibinfo{person}{Clara Isakowitsch}.} \bibinfo{year}{2022}\natexlab{}.
\newblock \showarticletitle{How augmented reality beauty filters can affect self-perception}. In \bibinfo{booktitle}{\emph{Irish Conference on Artificial Intelligence and Cognitive Science}}. Springer, \bibinfo{pages}{239--250}.
\newblock


\bibitem[LAURETIS(1987)]%
        {de1987technologies}
\bibfield{author}{\bibinfo{person}{TERESA~DE LAURETIS}.} \bibinfo{year}{1987}\natexlab{}.
\newblock \bibinfo{booktitle}{\emph{Technologies of Gender: Essays on Theory, Film, and Fiction}}.
\newblock \bibinfo{publisher}{Indiana University Press}.
\newblock
\showISBNx{9780253358530}
\urldef\tempurl%
\url{http://www.jstor.org/stable/j.ctt16gzmbr}
\showURL{%
\tempurl}


\bibitem[Leeat et~al\mbox{.}(2019)]%
        {Leeat_Shnabel_Glick_2019}
\bibfield{author}{\bibinfo{person}{R.~Z. Leeat}, \bibinfo{person}{N. Shnabel}, {and} \bibinfo{person}{P. Glick}.} \bibinfo{year}{2019}\natexlab{}.
\newblock \showarticletitle{The “prescriptive beauty norm” reflects a desire to enhance gender hierarchy and contributes to social policing of women and employment discrimination practices known as the “beauty tax.”}.
\newblock \bibinfo{journal}{\emph{Journal of Personality and Social Psychology}} (\bibinfo{year}{2019}).
\newblock
\newblock
\shownote{Available at: Harvard Kennedy School | Gender Actional Portal}.


\bibitem[Ma et~al\mbox{.}(2015)]%
        {ma2015cfd}
\bibfield{author}{\bibinfo{person}{D.~S. Ma}, \bibinfo{person}{J. Correll}, {and} \bibinfo{person}{B. Wittenbrink}.} \bibinfo{year}{2015}\natexlab{}.
\newblock \showarticletitle{The Chicago Face Database: A Free Stimulus Set of Faces and Norming Data}.
\newblock \bibinfo{journal}{\emph{Behavior Research Methods}}  \bibinfo{volume}{47} (\bibinfo{year}{2015}), \bibinfo{pages}{1122--1135}.
\newblock
\urldef\tempurl%
\url{https://doi.org/10.3758/s13428-014-0532-5}
\showDOI{\tempurl}


\bibitem[Madsen(2014)]%
        {Madsen2014}
\bibfield{author}{\bibinfo{person}{Ole~Jacob Madsen}.} \bibinfo{year}{2014}\natexlab{}.
\newblock \bibinfo{booktitle}{\emph{Governmentality}}.
\newblock \bibinfo{publisher}{Springer New York}, \bibinfo{address}{New York, NY}, \bibinfo{pages}{814--816}.
\newblock
\showISBNx{978-1-4614-5583-7}
\urldef\tempurl%
\url{https://doi.org/10.1007/978-1-4614-5583-7_126}
\showDOI{\tempurl}


\bibitem[McRobbie(2004)]%
        {mcrobbie2004post}
\bibfield{author}{\bibinfo{person}{Angela McRobbie}.} \bibinfo{year}{2004}\natexlab{}.
\newblock \showarticletitle{Post-feminism and popular culture}.
\newblock \bibinfo{journal}{\emph{Feminist media studies}} \bibinfo{volume}{4}, \bibinfo{number}{3} (\bibinfo{year}{2004}), \bibinfo{pages}{255--264}.
\newblock


\bibitem[Riccio et~al\mbox{.}(2024)]%
        {riccio2024mirror}
\bibfield{author}{\bibinfo{person}{Piera Riccio}, \bibinfo{person}{Julien Colin}, \bibinfo{person}{Shirley Ogolla}, {and} \bibinfo{person}{Nuria Oliver}.} \bibinfo{year}{2024}\natexlab{}.
\newblock \showarticletitle{Mirror, Mirror on the Wall, Who Is the Whitest of All? Racial Biases in Social Media Beauty Filters}.
\newblock \bibinfo{journal}{\emph{Social Media+ Society}} \bibinfo{volume}{10}, \bibinfo{number}{2} (\bibinfo{year}{2024}), \bibinfo{pages}{20563051241239295}.
\newblock


\bibitem[Riccio et~al\mbox{.}(2022)]%
        {riccio2022openfilter}
\bibfield{author}{\bibinfo{person}{Piera Riccio}, \bibinfo{person}{Bill Psomas}, \bibinfo{person}{Francesco Galati}, \bibinfo{person}{Francisco Escolano}, \bibinfo{person}{Thomas Hofmann}, {and} \bibinfo{person}{Nuria Oliver}.} \bibinfo{year}{2022}\natexlab{}.
\newblock \showarticletitle{OpenFilter: a framework to democratize research access to social media AR filters}.
\newblock \bibinfo{journal}{\emph{Advances in Neural Information Processing Systems}}  \bibinfo{volume}{35} (\bibinfo{year}{2022}), \bibinfo{pages}{12491--12503}.
\newblock


\bibitem[Ruggeri(2023)]%
        {Ruggeri_2023}
\bibfield{author}{\bibinfo{person}{A. Ruggeri}.} \bibinfo{year}{2023}\natexlab{}.
\newblock \bibinfo{booktitle}{\emph{The problems with TikTok’s controversial “beauty filters”}}.
\newblock BBC Online.
\newblock
\urldef\tempurl%
\url{https://tinyl.io/BjRX}
\showURL{%
\tempurl}
\newblock
\shownote{[Accessed: Nov 2024]}.


\bibitem[Spiel(2021)]%
        {spiel2021bodies}
\bibfield{author}{\bibinfo{person}{Katta Spiel}.} \bibinfo{year}{2021}\natexlab{}.
\newblock \showarticletitle{The bodies of tei--investigating norms and assumptions in the design of embodied interaction}. In \bibinfo{booktitle}{\emph{Proceedings of the Fifteenth International Conference on Tangible, Embedded, and Embodied Interaction}}. \bibinfo{pages}{1--19}.
\newblock


\bibitem[Winch(2015)]%
        {winch2015brand}
\bibfield{author}{\bibinfo{person}{A. Winch}.} \bibinfo{year}{2015}\natexlab{}.
\newblock \showarticletitle{Brand intimacy, female friendship and digital surveillance networks}.
\newblock \bibinfo{journal}{\emph{New Formations}} \bibinfo{volume}{84}, \bibinfo{number}{84-85} (\bibinfo{year}{2015}), \bibinfo{pages}{228--245}.
\newblock


\bibitem[Wissinger(2015)]%
        {wissinger2015year}
\bibfield{author}{\bibinfo{person}{Elizabeth Wissinger}.} \bibinfo{year}{2015}\natexlab{}.
\newblock \bibinfo{booktitle}{\emph{This year's model: Fashion, media, and the making of glamour}}.
\newblock \bibinfo{publisher}{NYU Press}, \bibinfo{address}{New York, USA}.
\newblock


\bibitem[Zuboff(2023)]%
        {zuboff2023age}
\bibfield{author}{\bibinfo{person}{Shoshana Zuboff}.} \bibinfo{year}{2023}\natexlab{}.
\newblock \showarticletitle{The age of surveillance capitalism}.
\newblock In \bibinfo{booktitle}{\emph{Social theory re-wired}}. \bibinfo{publisher}{Routledge}, \bibinfo{pages}{203--213}.
\newblock


\end{thebibliography}






\end{document}